\documentclass{elsart}
\usepackage{epsfig}
\usepackage{graphics}
\journal{Journal of Molecular Structure}
\begin{document}
\begin{frontmatter}

\title{Stability of the Liquid Phase in Colloidal Electrolytes}

\author{Jos\'e B. Caballero},
\author{Antonio M. Puertas}\ead{apuertas@ual.es}
\address{Group of Complex Fluids Physics, Department of Applied Physics, University of Almeria, 04120 Almeria, Andaluc\'{\i}a, Spain}
\date{\today}

\begin{abstract}
The equilibrium phase diagram of a 1:1 symmetrical mixture composed of oppositely charged colloids is calculated using Monte Carlo simulations. We model the system by the DLVO effective interaction potential. The phase diagram is similar to that of its atomic analog (the ionic fluid), where a liquid-gas first order transition emerges in the low $T-\rho$ regions being stable with respect to crystallization. As in the ionic fluids, we have found two different crystals: at high $T$ the fluid crystallizes in a FCC lattice, whereas at low $T$, the liquid coexists with a BCC crystal. The region of gas-liquid stability is observed to be narrower as the interaction range is diminished.
\end{abstract}

\end{frontmatter}

\section{Introduction}

The application of statistical mechanics concepts from atomic systems has boosted the understanding of more complex systems in the last decade, in particular of phase diagram of macromolecules in solution; i.e. the equilibrium phases arise from the competition of the energetic terms and the entropic ones \cite{anderson02}. Taking colloids as most simple macromolecules, without internal degrees of freedom, they still present some advantages: first, the interaction between the particles are easily tailored by external parameters (such as addition of salt or polymers), and secondly, the much larger length and time scales in colloids make experiments easier to handle. Additionally, in mixtures of colloids different interactions between provoke correlations that may introduce new phases.

For monocomponent colloidal systems, the phase diagrams are quite well understood, although not yet completely \cite{anderson02}. An attraction term in the interaction potential induces fluid-fluid coexistence, whose critical point moves to lower temperature as the range of the interaction is decreased. At a certain (short) range, the fluid-fluid coexistence becomes metastable with respect to crystallization \cite{anderson02,hagen94,lekerkerker92,ilett95}. The crystal phases are also affected, appearing an isostructural FCC-FCC transition as the range of the attraction narrows.

Recently, we studied the colloidal analog of the ionic fluid \cite{caballero04} (Restricted Primitive Model, RPM), i.e. a symmetrical mixture of oppositely charged colloids. As in the RPM, the colloidal mixture presents a gas-liquid separation in the low $T$ -- low $\rho$ regions where high correlations between oppositely charged particles are found \cite{fisher93}. Furthermore, the critical temperature evolves non-monotically as the range of the interaction (salt concentration in the medium) decreases \cite{cond-mat}. Experimental works with this system have focused on the crystal phases of such colloidal mixtures \cite{bartlett05,leunissen05}. Due to the high correlation between opposite colloids, different superlattice crystals are found, with cesium chloride structure for symmetrical systems (charge and size) at high attraction strengths and disordered-FCC for low ones, when the charge correlations are weak. Indeed, these structures are observed in the RPM \cite{bresme00}, and we seek them in the simulations in this work.

In this paper, our aim is to study the stability of the liquid for several interaction ranges, i.e. whether the freezing line crosses the liquid branch. We need for this to determine accurately the gas-liquid coexistence line, focusing on the finite size effects (on passing, we also show that this transition belongs to the 3D Ising criticallity class). On the other hand, the freezing line is estimated by melting a crystallyte, whose structure is also studied. The results indicate the existence of a stable liquid in a narrow region of temperatures for all of the ranges studied, and that the crystals have CsCl structure, in agreement with experimental results and with the RPM.

The paper is organized as follows: in Section II, we present the model and the simulations details are explained. Section III is divided in three subsections, where we study the liquid-gas transition in detail, the structure of the crystals, finally, the freezing line and the stability of the liquid-gas transition. We conclude in Section IV presenting the main conclusions.

\section{Model and Simulations}

\subsection{Model}

Our system is composed of a 1:1 binary mixture of $N$ spherical colloidal particles of equal diameter; $N/2$ bearing a surface potential $+\phi$ and $N/2$ with $-\phi$. The mixture is immersed in a continuous medium characterized by its dielectric constant, $\varepsilon$, in presence of an electrolyte. The electrostatic interactions are modeled using the well-known DLVO effective potential between the colloidal particles \cite{dlvo} and the dispersion forces are not considered. Thus the total potential is given by:  

\begin{equation} U(r)=\left\{ \begin{array}{ll}
\infty ,           & \hspace{0.5cm}  r \leq \sigma \\ [3mm]
 \pi \varepsilon \sigma \phi_1 \phi_2 \cdot \exp \left\{- \kappa (r - \sigma )\right\} , & \hspace{0.5cm}  r>\sigma 
\end{array} \label{pot_eq} \right. 
\end{equation}

\noindent where $\sigma$ is the hard core diameter of the particles, $\phi_1$ and $\phi_2$ are the surface potentials, and $\kappa$ is the inverse Debye length, which depends on the ionic concentration. If the charges of the particles are similar, the interaction in (1) is repulsive, whereas an attraction results if the particles carry opposite surface charges. This system is, thus, the colloidal analogue of the widely studied Restricted Primitive Model (RPM). This is obviously an approximation to the whole problem, where the ions are not simulated. However, the global problem, where the small ions are considered, requires a big amount of computational time and the coexistece lines cannot be easily obtained \cite{linse05}. The results here are, thus, a first approximation to the real problem.

Hereafter, reduced units will be used: $\sigma=1$, $U^{*}=U/(\pi\varepsilon\sigma\phi^{2})$ and so  $T^{*}=k_{B}T/(\pi\varepsilon\sigma\phi^{2})$ defines the reduced temperature, $\mu^{*}=\mu-k_{B}T\log\Lambda^{3}$ ($\Lambda$ is the thermal de Broglie wavelength)  is the reduced chemical potential and the density is defined as $\rho^{*}=N\sigma^{3}/V$ (where $N$ is the number of particles, and $V$ is the volume of the system). 

\subsection{Finite Size Scaling of Bruce and Wilding}

In the neighborhood of the critical point, the coexistence curve obtained from simulations depends strongly on the system size due to the long range of the correlations. This fact can be used to locate the critical point using the approach developed by Bruce and Wilding (BW) \cite{bruce92}. This method is based on the asymmetry of the density distribution, reflecting the absence of particle-hole symmetry in off-lattice models. Thus, the order parameter of the transition is not the density, but a mixture between the density and the energy: $M \sim \rho+su$ (where $s$ is a system-dependent field mixing parameter). Precisely at criticality, the distribution of $M$ in a large enough system, with box size $L$, takes on the universal form: $P_L(M)=P^*(x=a_L(M-<M>))$, where $P^{*}$ is an universal distribution function for each universality class, $<M>$ is evaluated at critical conditions and $a_{L} \sim L^{\beta/\nu}$. Thus, using GCMC simulations, critical parameters for different box sizes are calculated, and applying the corresponding scaling laws, the critical parameters for an infinite system can be obtained ($T_c(L)-T_c(\infty) \sim L^{-(\theta-1)/\nu}$, where $\nu$ is the critical exponent associated to the correlation length and $\theta$ is the universal correction to scaling exponent). This method can be improved including the pressure as a scaling field, but for Ising-type systems this contribution is usually negligible \cite{fisher04}. Since our model presents a short-ranged interactions, 3D Ising criticality is expected.

\subsection{Freezing Line}

When a solid phase is involved, techniques with insertion and removal of particles are highly inefficient since the insertion of a particle in a high density phase is strongly restricted. Hence, to locate the freezing line we carried out simulations in the canonical ensemble at high density. A system partially crystallized at the desired temperature and $\rho=0.90$ is then expanded to lower density, by small steps ($\delta \rho=0.01$) followed by an equilibration period. The crystallization degree is measured after equilibration for every density, by means of the global order parameter \cite{steinhardt83,wolde96}:

\begin{equation}
Q_{6} = \left( \frac{4 \pi}{13}\cdot \sum_{m=-6}^{6} \mid <Q_{6m}(\vec{r_{ij}})>\mid^{2} \right)^{1/2}
\end{equation}

\noindent where the angular brackets denote an ensemble average. $Q_{6m}(\vec{r_{ij}})$ is calculated for the neighbors $j$ of particle $i$: a neighbor is one particle within a given distance $r_{q}=1.5$ from $i$. $\vec{r_{ij}}$ is the vector which joins two neighboring particles and $Q_{6m}$ is related to the spherical harmonics: 
\begin{equation}
Q_{6m}(r) = Y_{6m} ( \theta (\vec{r_{ij}}), \phi (\vec{r_{ij}}))
\end{equation}

\noindent Note that $Q_{6}$ is invariant with respect to the reference coordinate system chosen. This global order parameter is close to zero in the fluid phase and has a non zero value when any degree of crystallization appears in the sample \cite{steinhardt83}. The freezing line is then located as the lowest density with a noticiable degree of crystallization.

\subsection{Simulation Details}

We firstly study the gas-liquid coexistence using Gibbs Ensemble Monte Carlo (GEMC) simulations \cite{frenkel}. In the GEMC technique, two independent boxes are hold at the same temperature, pressure and chemical potential by allowing exchanges of particles and volume between the boxes, but keeping the total volume and particle number constant throughout the simulation. We compare GEMC results using $N=2000$ particles and $N=432$ to study the effects of finite size on the coexistence lines and the critical points. 

Finite Size Scaling, based on the analysis of BW, have been performed by means of Grand Canonical Monte Carlo (GCMC) simulations aided with histogram reweighting techniques. Phase diagram with $L^*=10$ were obtained for $\kappa\sigma=6$ , $\kappa\sigma=10$ and $\kappa\sigma=15$. Grand Canonical Simulations comprised $25\cdot10^{6}$ steps for $L^*=8$ and $L^*=9$ and $75\cdot10^{6}$ steps for $L^*=10$ and $L^*=12$, each step consisted of $2·<N>$ attempts to insert or remove a random particle. 

To estimate the freezing line we use standard NVT Monte Carlo simulations with $N=1024$ particles. The particles are enclosed in a cubic box with periodic boundary conditions.

\section{Results and Discussion}

As we have discussed in previous papers \cite{caballero04,cond-mat}, this model undergoes a gas-liquid transition in the low $T$-low $\rho$ region; where high correlations between oppositely charged colloids arise, as in the RPM \cite{fisher93}. The critical temperature presents a non monotonic behaviour with the interaction range, what can be rationalized considering charge correlations \cite{cond-mat}. On average, each charged colloid is sorrounded by a layer of oppositely charged particles, followed by a layer of similarly charged colloids and so on. Therefore, when the salt concentration is increased, the attractive term is shortly screened (because the opposite charged particles are in contact); contrary, the repulsive contributions are strongly screened since longer distances separate unequal colloids. The system, thus, gains energy upon increasing $\kappa\sigma$, resulting in an increase of the critical temperature (opposite to the behaviour of monocomponent systems). At high enough $\kappa$, the repulsive interactions are completely screened, $T_c$ decreases as in a monocomponent system, since only the first layer of particles (with opposite charge) interacts, leaving a maximum at $\kappa\sigma \approx10$. An island of phase separation (gas-liquid) is then predicted for this model. 

As presented in the introduction, the aim of this paper is to study whether the triple temperature is lower than the critical one around $\kappa\sigma$ values which are close to the maximum $T_c$. We will study thus the liquid-gas transition, and then the freezing line. For the latter, we need first to form stable crystals at different temperatures.

\subsection{Liquid-gas transition}

\begin{figure}
\begin{center}
\psfig{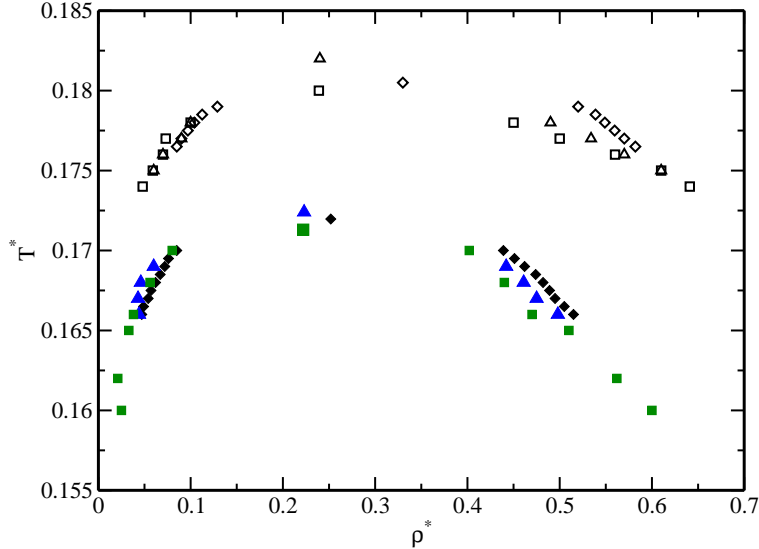}
\end{center}
\caption{\small \it
Gas-liquid transition for $\kappa\sigma=6$ (filled symbols) and $\kappa\sigma=10$ (open symbols):  GEMC results with $N=432$ (squares) and $N=2000$ (triangles) and GCMC results (diamods). The critical point in the GEMC simulations is calculated by means of the law of rectilinear diameters.}
\label{diagram}
\end{figure}

GEMC and GCMC results for the liquid-gas transition at $\kappa\sigma=6$ and $\kappa\sigma=10$ are presented in Fig. \ref{diagram}. Note that, contrary to monocomponent systems, the critical point is at higher temperature for the system with the shorter interaction range. To check for finite size effects, we compare in this figure simulations with $N=432$ and $N=2000$ particles (GEMC), and with those from GCMC simulations aided with reweighting techniques with $L^{*}=10$. Although the different estimations of the critical temperautures coincide quite well, bigger differences for the critical densities are noticed, in agreement with previous comparisons for the RPM and other models \cite{orkoulas99,pana02}. The coexistence curves coincide for the three cases far away from the critical point, and differ slightly only close to the critical point, where the correlation length becomes similar to (or even larger than) the simulation box.

\begin{figure}
\begin{center}
\psfig{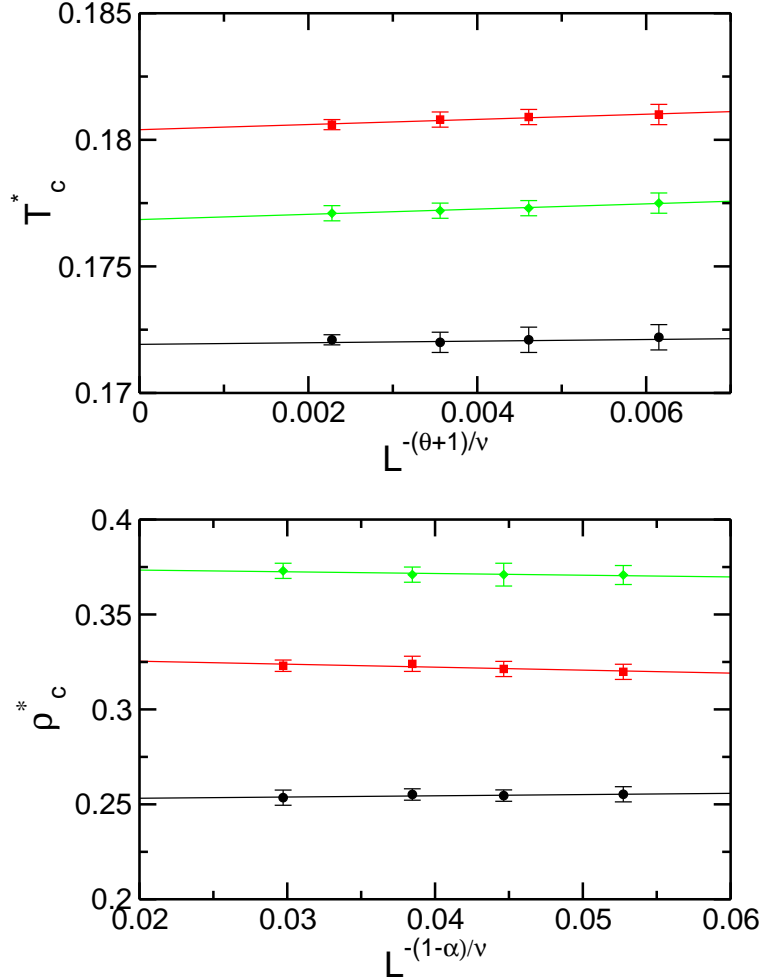}
\end{center}
\caption{\small \it
Critical temperature (upper panel) and density (lower panel) as a function of the system size, using the corresponding scaling laws and the exponents from the 3D-Ising universality class. Black circles: $\kappa\sigma=6$, red squares: $\kappa\sigma=10$ and green diamonds: $\kappa\sigma=15$.}
\label{finite_size}
\end{figure}

The behaviour of the critical parameters with the system size is studied in Fig. \ref{finite_size} by means of GCMC simulations and using the corresponding scaling laws for the temperature (upper panel) and density (lower panel). Since the interactions are short ranged, the scaling exponents are taken from the 3D Ising model. This assumption is further supported by the analysis of the marginal distribution (see below). Both the critical temperature and density show only slight dependences on the system size, and the values extrapolated to an infinite system differ very little from the critical values for the finite systems presented.

In GCMC criticalicity is recognized when the marginal distribution of the simualtion coincides with that of the 3D Ising model, that is, the universality class must be assumed a priori. The best way of working would be to compare the distribution with different universal distributions, but for other models the critical distributions are not known. Therefore, we can only carry out a study of compatibility between our results and the 3D Ising universality class, which, nevertheless, is expected, due to the short range of the interactions.

\begin{figure}
\begin{center}
\psfig{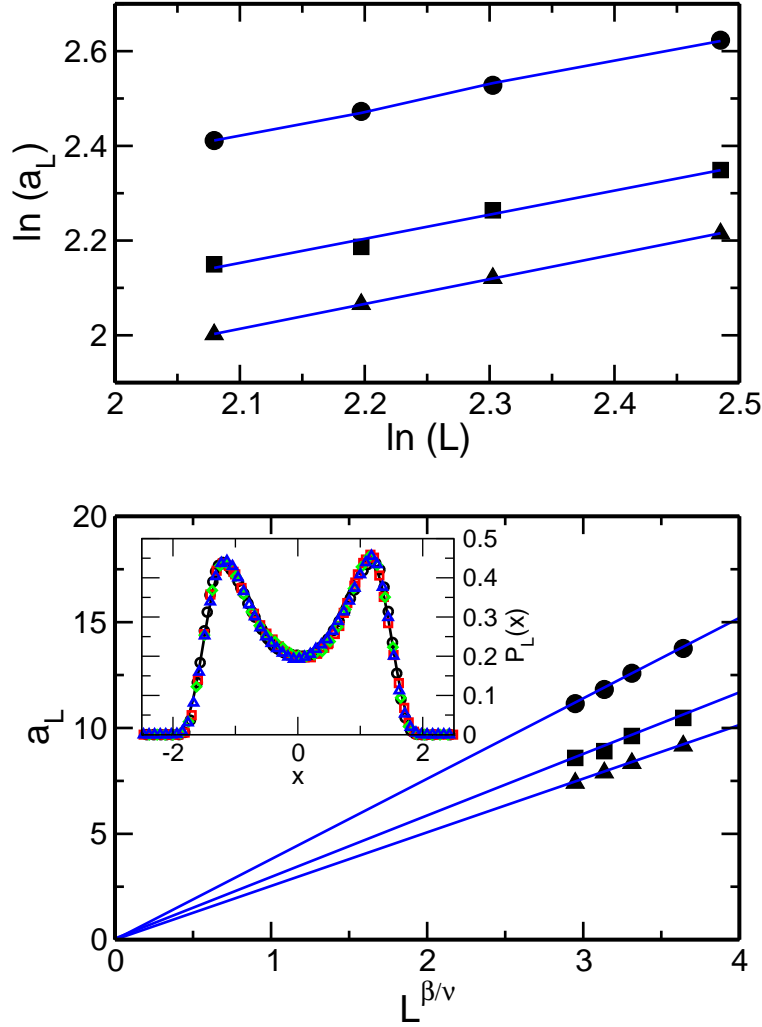}
\end{center}
\caption{\small \it
Critical behavior study for $\kappa \sigma=6$ (circles), $\kappa \sigma=10$ (squares) and $\kappa \sigma=15$ (diamods). Higher panel: estimation of the critical ratio $\beta / \nu$. Lower panel: $A_{M}(L)$ versus $L^{\beta / \nu}$. Blue lines are lineal fits for the numerical results. Inset: order parameter probability distribution $P_L(x)$ for $\kappa\sigma=10$ and box sizes: $L=8$ (circles), $L=9$ (squares), $L=10$ (diamonds) and $L=12$ (triangles); solid line for 3D Ising model.}
\label{ising}
\end{figure}

The marginal distribution function ($P_L(x)$) is mapped onto the 3D Ising distribution in the inset of Fig. \ref{ising} (lower panel), and the critical conditions are determined when the match between both distributions is achieved. Note that the distributions obtained from simulations match very well the universal Ising one. Additionally, using the relation $a_{L} \sim L^{\beta/ \nu}$ the ratio between the critical exponents $\beta$ and $\nu$ can be obtained. The values computed are similar to those of the Ising model ($\beta/\nu=0.518$) for the three ranges studied: $\beta/\nu=0.523(14), 0.509(23)$ and $0.503(20)$ for $\kappa\sigma=6, 10$ and $15$, respectively (upper panel in Fig. \ref{ising}).

To finish the study of the GCMC data, we follow the analysis which was performed by Camp and Patey \cite{camp01}. They showed that for systems with 3D Ising universality class, the relation $a_L=BL^{\beta/ \nu}$ should be fulfilled (where B is a system-dependent parameter). In the lower panel of Fig. \ref{ising}, the results for this analysis are plotted. Note that for the three $\kappa\sigma$ values studied the results can be fitted with straight lines crossing the origin, indicating again the compatibility between our results and the Ising criticalicity. Because we have always found compatibility between our GCMC results and the Ising universality class, we can state that the critical parameters and the coexistence curve from GCMC are expected to be more accurate than the GEMC ones. However, since we are interested in the stability of the liquid phase, if the triple temperature is much lower than $T_c$, we can indistincly use GCMC or GEMC results because far away from the critical point both coexistence curves coincide (Fig. \ref{diagram}).

\subsection{Structure of the crystal}

In the RPM, three different crystal phases have been reported \cite{bresme00}: disordered FCC at high temperatures, where the ions are randomly located; ordered BCC or CsCl in the low $T$ regions, with each ion in the center of a cubic box surrounded by eight opposite ions in the vertexes of the cubic box, which, upon compresion suffers a first order transition to an ordered FCC structure. Due to the analogies between the colloidal mixture and the RPM, similar phases are to be expected here. Since we only attempt to nail down the freezing line, we will not discuss here the possibility of the transition between CsCl and ordered-FCC structures.

\begin{figure}
\begin{center}
\psfig{file=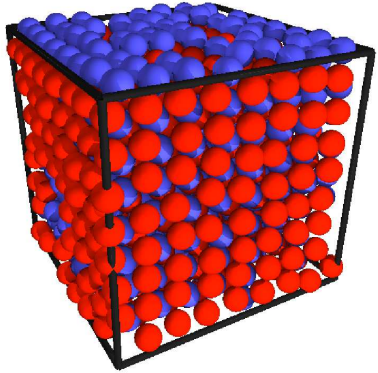,width=10cm}
\end{center}
\caption{\small \it
Snapshot of the system at $T^{*}=0.17$ and $\rho^{*}=0.90$ for $\kappa \sigma=6$, showing liquid-crystal coexistence. Note the ordering of the particles, every particle has eight neighbours with opposite charge. This systems yields $Q_6=0.41$}
\label{snapshot}
\end{figure}

\begin{figure}
\begin{center}
\psfig{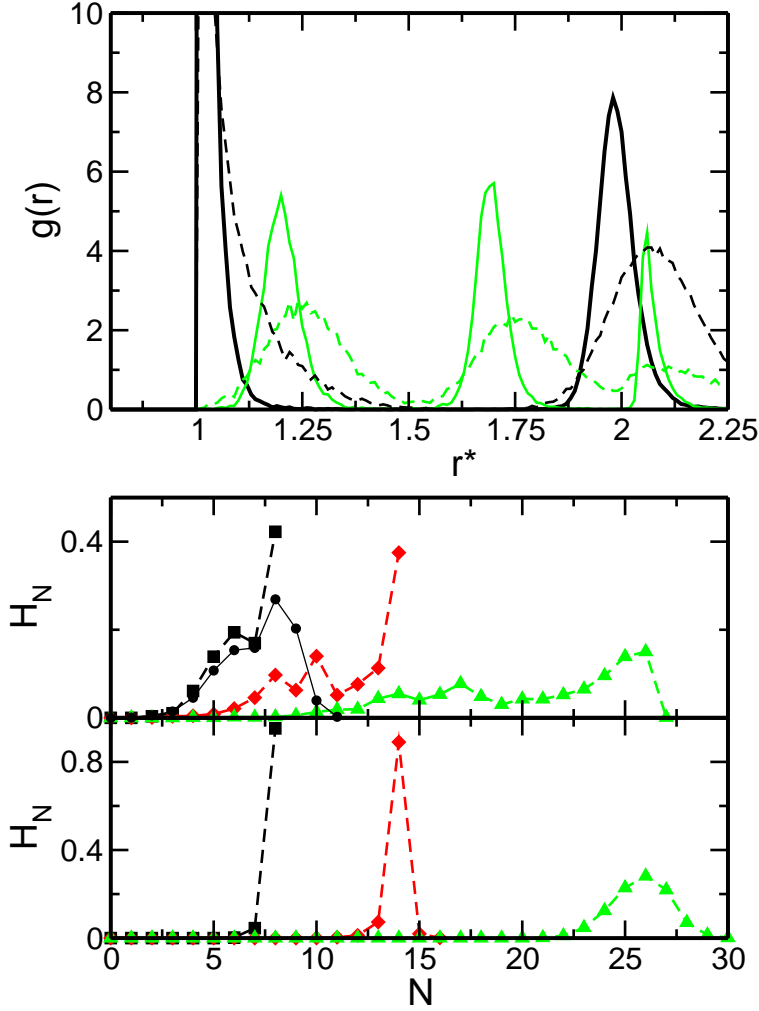}
\end{center}
\caption{\small \it
Upper panel: partial radial distribution functions for the system at $T^{*}=0.13$ and $\rho^{*}=0.30$ for $\kappa \sigma=6$ (continuous lines) and at $T^{*}=0.17$ and $\rho^{*}=0.90$ (broken lines). $g_{+-}(r)$ with black line and $g_{++}$ with green line. Lower panel: for both states, histograms of the number of particles inside a sphere centered in each particle with radius $R_C$: small cirlces for $R_{C}=1.15$ (first layer of particles); squares for $R_{C}=1.15$ considering only neighbors of oppositely charged; diamonds for $R_{C}=1.4$ (second layer of particles) and triangles for $R_{C}=1.85$ (third layer of particles).}
\label{bcc}
\end{figure}

In Fig. \ref{snapshot} we present a snapshot of the system at $T^{*}=0.17$ and $\rho^{*}=0.90$ for $\kappa\sigma=6$, which has very clearly crystallized into a CsCl structure. This temperature is very similar to the critical one for this range, which correspond to ``low temperatures'', in the comparison with the RPM results. This picture compares nicely with experimental ones for symmetrical mixtures of charged colloids presented recently in similar conditions \cite{bartlett05,leunissen05}. Similar crystallites were observed at other low temperatures and not-too-high density. At lower temperatures, however, the crystallite is in coexistence with a vapour phase, instead of a liquid phase, signalling the triple temperature. The system in the snapshot above and a system with $T^{*}=0.13$ and $\rho^{*}=0.3$ (gas-crystal coexistence) are analysed in Fig. \ref{bcc}. 

In the upper panel, the partial radial distribution functions are shown (black line for $g_{+-}(r)$ and green line for $g_{++}(r)$). Note that the system is composed of alternating layers of opposite charges, similarly to the structure of the liquid phase. Noteworthly, the state with the lowest temperature yields narrower peaks, since the vapor is very dilute and it does not contribute noticeably to $g(r)$, whereas the liquid smears out the crystal peaks at higher temperatures.

\begin{figure}
\begin{center}
\psfig{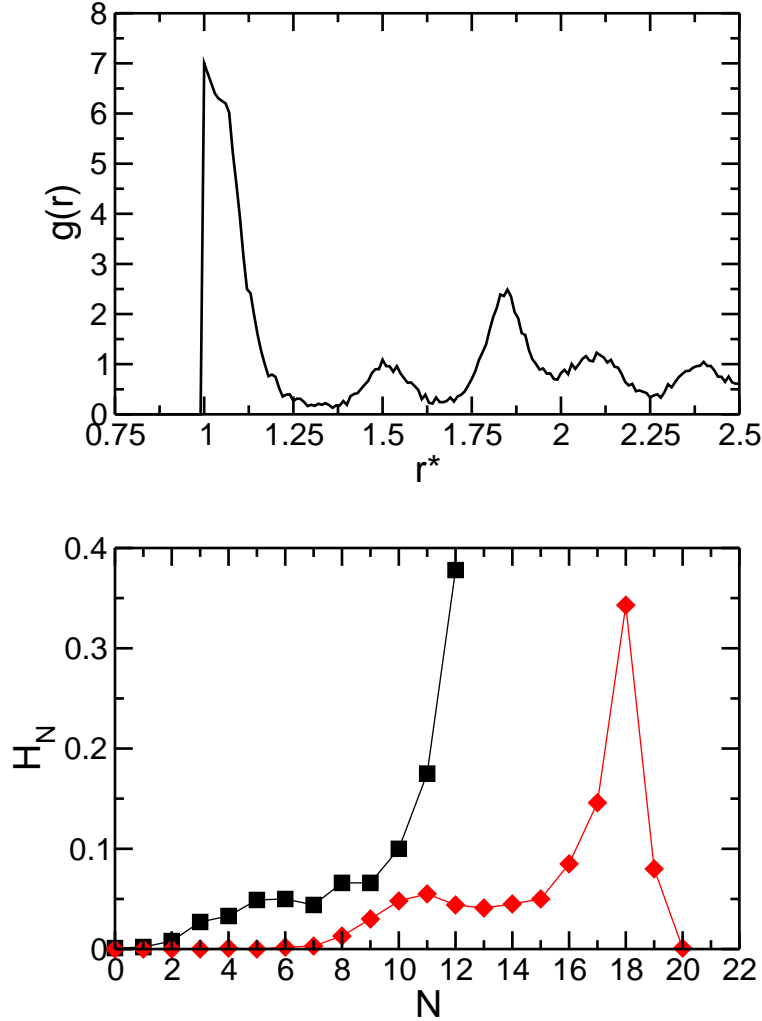}
\end{center}
\caption{\small \it
Higher panel: total radial distribution function at $T^{*}=2$ and $\rho^{*}=0.95$ for $\kappa \sigma=6$ (fluid-crystal coexistence). Lower panel: as in Fig. \ref{bcc}: squares for $R_{C}=1.25$ and diamonds for $R_{C}=1.6$.}
\label{fcc}
\end{figure}

The distribution of the number of particles inside a sphere centered in each particle with radius $R_C$ is presented for both states in the lower panel of Fig. \ref{bcc} (the central particle is not counted). When $R_C=1.15$, only nearest neighbours are considered, i.e. the number of particles inside the first peak in g(r). Since the first and second layers overlap (see the upper panel), some colloids are surrounded by $10$ particles. Nevertheless, when only particles with opposite sign are exclusively taken into account, the maximum number of nearest neighbours is $8$. That is, each sphere is surrounded at most by eight oppositely charged particles. Taking $R_C=1.4$, the particles inside the first and second layers are being counted. Now the maximum number of particles inside such a sphere is $14$. Finally, setting $R_C=1.85$, we consider up to the third layer of particles and $24$ particles were found. All these data are compatible with a BCC strucutre (as in the CsCl), but not with the FCC structure. As in the RPM, in the low $T$ regions the fluid (liquid or vapour) coexists with an ordered-BCC crystal. The stability of such crystals is due to the energetic terms which favor the formation of BCC-like crystals because the number of contacts with particles of opposite sign is higher than in the FCC structure, even though the entropic contribution aims for the more compact FCC structure.

At higher temperatures, the results obtained are completely different. Fig. \ref{fcc} shows the radial distribution function and the distribution of the number of particles inside spheres with different radii ($T^{*}=2$ and $\rho^{*}=0.95$). The partial structure is now random, that is, structuration in layers was not found. The distribution of particles is (lower panel in Fig. \ref{fcc}): $12$ particles in the first layer and $6$ particles in the second one, corresponding to the disordered-FCC crystal. Both structures shown here fully agree with the recent experimental results on the same system \cite{bartlett05,leunissen05} and with the expectations from the RPM \cite{bresme00}.

\subsection{Stability of the liquid}

\begin{figure}
\begin{center}
\psfig{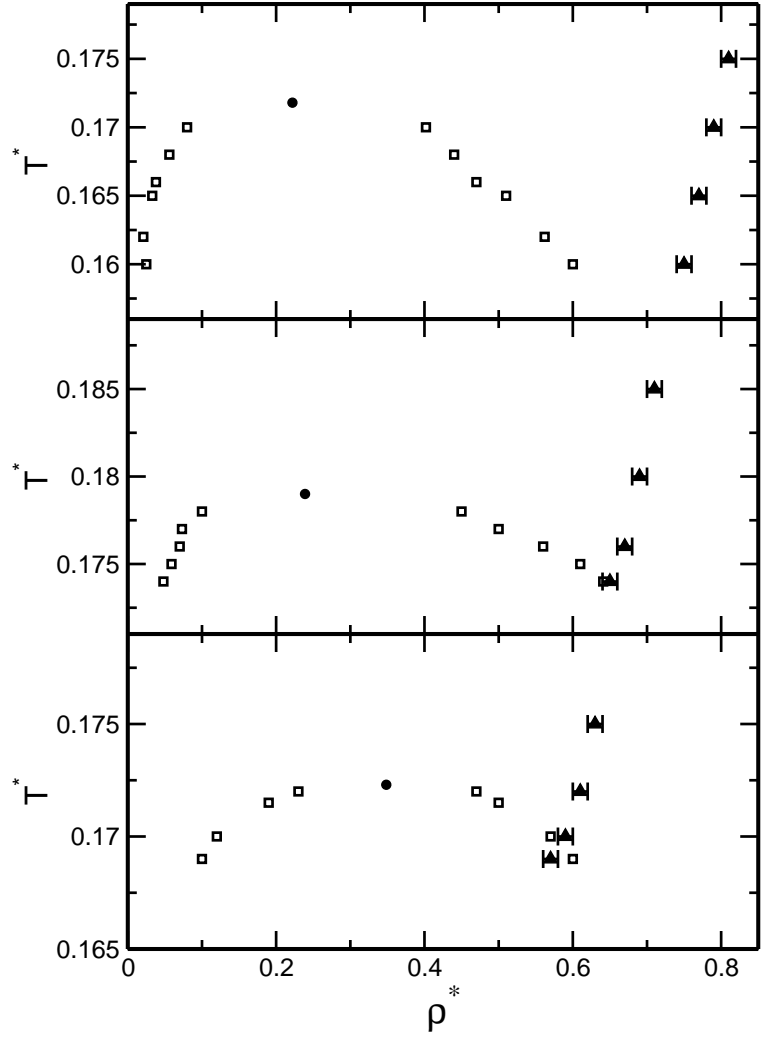}
\end{center}
\caption{\small \it
Gas-liquid coexistence curves from GEMC (open circles and filled circles for the critical points) and freezing line (filled triangles) for $\kappa\sigma=6$ (higher panel), $\kappa\sigma=10$ (middle panel) and $\kappa\sigma=15$ (lower panel). Note that the $T*$-scales differ for different panels.}
\label{complete}
\end{figure}

The freezing line can be determined using the procedure described in Section II: we take a spontaneuosly formed crystal seed at high enough density and decrease the density, with small steps, down to the point where the crystal is completely melted. Throughout this path, the global orientation parameter $Q_6$ was used as the order parameter. Fig. \ref{complete} shows the numerical estimations of the freezing line together with the gas-liquid coexistence points for three different $\kappa\sigma$ values ($\kappa\sigma=6$, where the critical temperature grows with $\kappa$; $\kappa\sigma=10$, $T_c$ reaches its maximum value and $\kappa\sigma=20$, $T_c$ behaves as in monocomponent systems). For these three values studied here, the liquid phase is stable with respect crystallization as we can observe in Fig. \ref{complete} in a narrow window of temperaures. Interestingly, the freezing line behaves monotonically with $\kappa\sigma$, moving to lower density (or higher temperature), contrary to the non-monotonous behaviour of the liquid-gas transition.

\begin{figure}
\begin{center}
\psfig{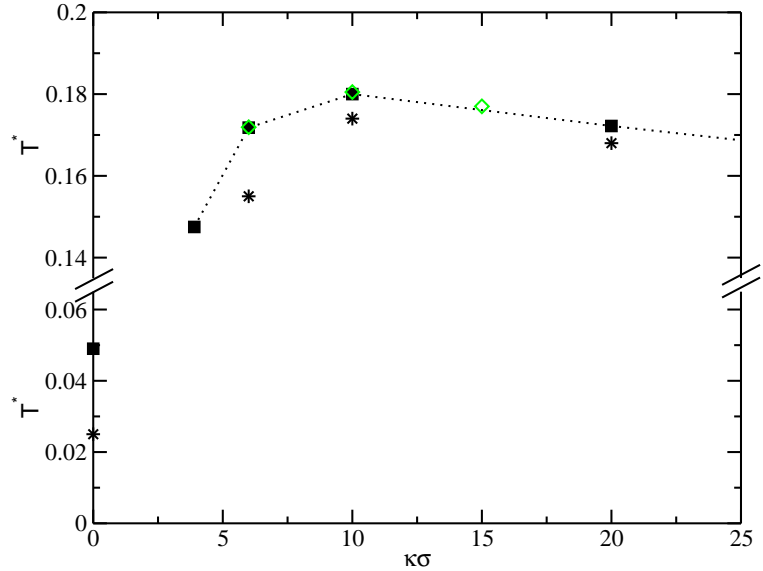}
\end{center}
\caption{\small \it
Squares: critical temperatures of the colloidal mixture from GEMC (triangles from GCMC) and the ionic fluid \cite{pana02}. Stars: triple temperatures (estimated value for $\kappa\sigma=6$). The symbols at $\kappa\sigma=0$ mark the RPM values \cite{bresme00}.}
\label{triple}
\end{figure}

The triple point is set by the crossing between the freezing line and the liquid branch. Fig. \ref{triple} plots the critical and triple temperatures for the colloidal system and its atomic analog, the ionic fluid. Note that the stable gas-liquid gap decreases as $\kappa\sigma$ increases. Extrapolating the present results, the triple temperature is expected to be equal to the critical one around $\kappa\sigma \sim 25$, thus implying that the liquid-gas transition is metastable with respect to crystallization for shorter interaction ranges. This prediction cannot be proved straight away because more sophisticated algorithms should be used to sample correctly the whole space phase for systems with such short-ranged interactions \cite{frenkel05}.

\section{Conclusions}

We have studied the stability of the liquid phases for the colloidal analog of the ionic fluid, where the system was modeled by the effective DLVO interaction potential. This system presents richer behavior than the ionic one \cite{caballero04,cond-mat,bartlett05,leunissen05}, because the range of the interaction can be tuned and the electroneutrality is imposed by construction. The liquid phases in these mixtures are stable for the range of the interaction studied, but the existence of the liquid is restricted to narrower T-regions as the interaction range is decreased.

BCC crystals have been found at low $T$, but at higher temperatures, FCC crystals appear, when the correlations between oppositely colloidal particles are not important. These results confirm that the phase diagram of a symmetrical colloidal mixture presents similar aspect to that of the RPM (while the liquid is stable). Even, experimental studies support such a statement \cite{bartlett05,leunissen05}.

\ack

This work is financially supported by the MCyT under projects no. MAT2004-03581 and MAT2003-03051-CO3-01. The authors thank Prof. L.F. Rull and Dr. J.M. Romero-Enrique for useful discussions.

\end{document}